\documentclass{article}

\usepackage[numbers,compress,square]{natbib}
\usepackage[margin=1in]{geometry}

\usepackage[utf8]{inputenc} 
\usepackage[T1]{fontenc}    
\usepackage{hyperref}       
\usepackage{url}            
\usepackage{booktabs}       

\usepackage{amsfonts}       
\usepackage{nicefrac}       
\usepackage{microtype}      
\usepackage{xcolor}         

\usepackage{amsmath}
\usepackage{amsthm}
\usepackage{mathtools}
\usepackage{subcaption}
\usepackage{algorithm}
\usepackage[noend]{algpseudocode}
\usepackage{comment}
\usepackage{multirow}
\usepackage{graphicx}
\usepackage{wrapfig}
\usepackage{floatflt}

\setcitestyle{brackets=square}

\newtheorem{theorem}{Theorem}
\newtheorem{corollary}{Corollary}
\newtheorem{lemma}{Lemma}

\newtheorem{question}{Open Question}

\DeclareMathOperator{\err}{err}
\DeclareMathOperator{\ltt}{LTToep}
\DeclareMathOperator{\atan2}{atan2}
\def\mcount{M_{\textit{count}}}

\title{Private Continual Counting of Unbounded Streams}

\author{%
  Ben Jacobsen \\
  University of Wisconsin--Madison \\
  \texttt{bjacobsen3@wisc.edu}
  \and
  Kassem Fawaz \\
  University of Wisconsin--Madison \\
  \texttt{kfawaz@wisc.edu}
}

\begin{document}

\maketitle

\begin{abstract}
  We study the problem of differentially private continual counting in the unbounded setting where the input size $n$ is not known in advance. Current state-of-the-art algorithms based on optimal instantiations of the matrix mechanism~\citep{li_matrix_nodate} cannot be directly applied here because their privacy guarantees only hold when key parameters are tuned to $n$. Using the common `doubling trick' avoids knowledge of $n$ but leads to suboptimal and non-smooth error. We solve this problem by introducing novel matrix factorizations based on logarithmic perturbations of the function $\frac{1}{\sqrt{1-z}}$ studied in prior works, which may be of independent interest. The resulting algorithm has smooth error, and for any $\alpha > 0$ and $t\leq n$, it is able to privately estimate the sum of the first $t$ data points with $O(\log^{2}(t)(\log\log t)^{1+2\alpha})$ variance. It requires $O(t)$ space and amortized $O(\log t)$ time per round, compared to $O(\log(n)\log(t))$ variance, $O(n)$ space and $O(n \log n)$ pre-processing time for the nearly-optimal bounded-input algorithm of \citet{henzinger_almost_2023}. Empirically, we find that our algorithm's performance is also comparable to theirs in absolute terms: with optimized parameters, our variance is less than $2\times$ theirs for $t$ as large as $2^{32}$.
\end{abstract}

\begin{center}
\noindent\fbox{
    \parbox{0.9\textwidth}{\textbf{Note}: This version of the paper corrects a numerical integration error in the version published at NeurIPS 2025, which caused the implementations used in our empirical evaluation to satisfy a weaker-than-intended privacy guarantee. Section 5 has been updated with the corrected implementation and additional discussion on parameter selection.}
  }
\end{center}

\section{Introduction}

\textit{Differentially private counting under continual observation}~\citep{chan2011private,dwork_differential_nodate} refers to the problem of maintaining accurate running totals over streams of sensitive data.  It has attracted a great deal of attention in recent years~\citep{andersson_smooth_nodate,andersson_streaming_2024,andersson_continual_nodate,bienstock_dmm_2024,denisov2022improved,dvijotham_efficient_2024,fichtenberger2023constant,henzinger_almost_2023,henzinger2024unifying,mcmahan_inversion_2025} for its wide-ranging applications in optimization~\citep{choquette2023amplified,choquette2022multi,denisov2022improved,dwork2014analyze,kairouz2021practical,mcmahan2024hassle}, as well as online learning and the private analysis of streaming data more broadly~\citep{agarwal2017price,cardoso2022differentially, jain_differentially_nodate,smith_nearly_nodate,upadhyay2019sublinear}. It has also been used in many large-scale deployments of differential privacy, such as Google's private next-word prediction~\citep{mcmahan2022federated} and Smart Select models~\citep{hartmann2023distributed}. To a large extent, this flurry of activity has been prompted by recent algorithmic improvements in matrix factorizations for streaming data~\citep{denisov2022improved, fichtenberger2023constant,henzinger_almost_2023,li_matrix_nodate}, which have dramatically improved privacy/utility tradeoffs compared to classical approaches based on binary trees~\citep{chan2011private,dwork_differential_nodate,honaker2015efficient}.

One challenge when applying matrix-based algorithms to streaming settings is that they largely assume access to the input size $n$ as a parameter. While this is a very natural assumption in contexts like private model training, researchers from the beginning have also motivated their work with appeals to applications like public health~\citep{dwork_differential_nodate} where it is much less clear how the assumption could hold. In fact, it is known that difficulties related to unbounded inputs have influenced practical deployments of DP~\citep{amin2024practical}: famously, after Apple announced that it planned to use differential privacy to collect user data, \citet{tang2017privacy} discovered that their implementation achieved this by guaranteeing privacy only over a single day while allowing privacy loss per-user to grow without bound over time.

We consider the setting in which the input size $n$ is not known in advance. Instead, we are only given a promise that $n \leq N$ for some $N \leq +\infty$. Our goal is to design algorithms whose error and time complexity automatically scale with the true input size $n$ while guaranteeing privacy for any input with size up to $N$. 

Existing solutions to this problem all involve fundamental tradeoffs.
Tuning a bounded-input mechanism to the worst-case input size $N$ guarantees privacy, but leads to error that scales with $N$ even when $n \ll N$.
Lifting the bounded-input algorithm to the unbounded case with some form of doubling trick~\citep{chan2011private} avoids this issue while preserving performance asymptotically, but introduces lower-order error terms which can be significant at finite sample sizes. Using a doubling trick also leads to non-smooth growth in error over time, undermining one of the major selling points of matrix methods over binary trees~\citep{henzinger_almost_2023}. 

Our main contribution in this paper is a matrix-based algorithm that overcomes this dilemma: to the best of our knowledge, it is the first private counting algorithm which is simultaneously \textbf{smooth} and \textbf{unbounded} with \textbf{almost-optimal asymptotic error}. It is also practical to implement, with essentially the same complexity as the foundational algorithm of \citet{henzinger_almost_2023}, and we show empirically that it enjoys small error for realistic values of $n$ and $N$ and not just asymptotically.

\section{Background and Related Work}

\subsection{Background}
\paragraph{Continual (Binary) Counting.} In continual counting~\citep{dwork_differential_nodate, chan2011private}, the input consists of a stream of data $x_1, x_2,\ldots \in \mathcal{X}^n$. We assume that $n \leq N$ for some positive constant $N \leq +\infty$; $N$ is known in advance, but $n$ is not. At each time step $t$, the algorithm receives $x_t$ and then produces an estimate of the partial sum $S_t = \sum_{i=1}^t x_i$. For the sake of simplicity, we will assume that $\mathcal{X} = \{0, 1\}$, but our results are easy to generalize to $\mathcal{X} = \mathbb{R}^d$.

\paragraph{Differential Privacy under Continual Observation.} Define two (potentially unbounded) streams $x_1, x_2,\ldots$ and $x'_1, x'_2, \ldots$ to be \textit{neighboring} if there is at most one time step $i$ where $x_i \neq x'_i$. Let $X_t$ (resp. $X'_t$) denote the first $t$ data points in each stream. We say that a mechanism $\mathcal{M}$ satisfies $(\varepsilon, \delta_{priv})$-differential privacy (DP)~\citep{dwork2014algorithmic} in the continual observation model~\citep{dwork_differential_nodate} if for all neighboring streams $X,X'$, all $t \in \mathbb{Z}^+$, and all measurable events $S$, we have \( \mathbb{P}(\mathcal{M}(X_t) \in S) \leq e^{\varepsilon} \mathbb{P}(\mathcal{M}(X'_t) \in S) + \delta_{priv}\). Here, $\mathcal{M}(X)$ represents the full sequence of outputs produced by $\mathcal{M}$ on input $X$. This reduces to the standard definition when the streams are finite; the additional quantification over $t$ is necessary to make sense of the definition for unbounded streams. The mechanisms we study all use additive Gaussian noise, and so can readily be shown to satisfy other notions of indistinguishability as well, such as R\'enyi DP~\cite{mironov2017renyi}, zCDP~\cite{bun2016concentrated}, or Gaussian DP~\cite{dong2022gaussian}. By virtue of using Gaussian noise, they also automatically satisfy DP in the stronger \textit{adaptive} setting where $x_t$ is permitted to depend on earlier outputs, which is important for learning applications ~\citep{denisov2022improved}.

Two measures of utility are standard in the literature: the root mean squared additive error and the expected maximum additive error, defined respectively as the maximums over $X_n$ of:
\begin{align*}
    &\err^{\ell_2}_n(\mathcal{M}) := \sqrt{\frac{1}{n} \mathbb{E}_{\mathcal{M}} \big( \sum_{t \leq n} (\lVert S_t - \mathcal{M}(X)_t \rVert_2^2  \big)\big)}, 
    &\err^{\ell_\infty}_n(\mathcal{M}) := \mathbb{E}_{\mathcal{M}}  \big( \max_{t \leq n} ~\lVert S_t - \mathcal{M}(X)_t \rVert_\infty \big).
\end{align*}
When $\mathcal{M}$ uses additive Gaussian noise, both expectations and matching high-probability bounds can be derived from $\mathbb{V}(\mathcal{M}(X)_t)$, the (non-asymptotic) variance of the noise added as a function of time. This is the main metric we will report when comparing the performance of different algorithms.

We say that a mechanism is \textit{unbounded} if $\mathbb{V}(\mathcal{M}(X)_t) = O_N(1)$, where the big-Oh is understood as taking $N \to \infty$ for fixed $t$. We additionally say that a mechanism has \textit{smooth error} if $\mathbb{V}(\mathcal{M}(X)_t)$ is a smooth function of $t$. This is a desirable property because $\err^{\ell_\infty}_n$ depends on the maximum variance across all time steps, which can be significantly higher than the average variance when error is non-smooth~\citep{andersson_smooth_nodate}.

\subsection{Related Works}

\textit{Differential privacy under continual observation} was introduced by \citet{dwork_differential_nodate} and \citet{chan2011private}, who independently proposed the \textit{binary mechanism}. Conceptually, this algorithm computes partial sums with the help of a binary tree where leaf nodes contain private estimates of individual data points, and internal nodes contain independent estimates of the subsequences spanned by their children. 
The binary mechanism of \citet{dwork_differential_nodate} used Laplacian noise to satisfy $(\varepsilon,0)$-DP with $\err_t^{\ell^\infty} = O(\log^{2.5}(n))$, which was improved to $O(\log^{2.5}(t))$ by \citet{chan2011private}. This was later reduced to $O(\log(t) \sqrt{\log n})$ by \citet{jain_differentially_nodate} by using Gaussian noise instead.
A straightforward optimization of the algorithm follows from the observation that at a given time step, it is only necessary to store $\log_2(n)$ nodes of the tree in memory. Subsequent work by \citet{honaker2015efficient} additionally showed that it is possible to reduce the error by roughly a factor of 2 by making efficient use of multiple independent estimates for the same partial sum. 

Both the original binary mechanism and Honaker's variant have very uneven noise scale across time steps. Recently, \citet{andersson_smooth_nodate} proposed a smooth variant of the binary mechanism which equalizes the noise distribution across time steps, further improving $\err^{\ell_\infty}_n$ by a constant factor.

The binary mechanism can be seen as a special case of the general matrix method~\citep{li_matrix_nodate}. Given a dataset $x$ and a matrix $A = LR$, the matrix method estimates $Ax$ by $L(Rx + z) = Ax + Lz$. This satisfies DP when $z$ is a Gaussian scaled to $\lVert R \rVert_{1\to2}$, the maximum $\ell_2$ norm over the columns of $R$. The final error also scales with $\lVert L \rVert_{2\to\infty}$, the maximum $\ell_2$ norm over the rows of $L$, and so by choosing $L$ and $R$ carefully we can achieve error depending on the factorization norm $\gamma_2(A) = \min \lbrace \lVert L \rVert_{2\to\infty} \lVert R \rVert_{1\to 2} \mid LR = A \rbrace$. This is often much more accurate than calibrating $z$ to $A$ directly. In the case of continual counting, $A = \mcount$, the all-ones lower-triangular matrix.

This matrix-based approach has rapidly become the standard in the field. \citet{edmonds_power_2019} proved a lower bound of $\Omega(\log n)$ on the per-time-step $\ell_\infty$ error of continual counting by lower-bounding $\gamma_2(\mcount)$.  \citet{denisov2022improved} approached the problem of finding optimal decompositions as a convex optimization problem, building on earlier work in the offline setting ~\citep{yuan2016optimal}. They proved that, for lower-triangular matrices like $\mcount$, it is sufficient to consider only decompositions where $L$ and $R$ are both lower-triangular. This is a particularly nice result because it implies there is no extra cost for using the matrix method in streaming settings. Subsequently, \citet{fichtenberger2023constant} provided an explicit factorization $L = R = M_{\textit{count}}^{1/2}$ and studied its $\err^{\ell_\infty}_n$ error, and \citet{henzinger_almost_2023} proved that the same decomposition is simultaneously nearly optimal in terms of $\err^{\ell_2}_n$. More recent work has investigated various computational modifications of this factorization, trading a small amount of accuracy for efficiency by approximating the optimal decomposition in more compact ways, e.g. through finite recurrences~\citep{dvijotham_efficient_2024, andersson_space_2025} or by binning together similar values~\citep{andersson_streaming_2024}.

A distinct line of work has investigated the unbounded case. In their seminal work, \citet{chan2011private} proposed a `hybrid' mechanism which uses a variant of the classic doubling trick to match the asymptotic $\err_n^{\ell_\infty}$ rate of the binary mechanism when $n$ is unknown. Their technique is generic and can be used to `lift' any fixed-size algorithm into one that works on unbounded streams with the help of a (potentially inefficient) unbounded algorithm --- in \autoref{sec:extensions} we investigate a version of this mechanism instantiated with \autoref{alg:log_matrix}. An alternative strategy for handling unbounded inputs is to only permit queries based on sliding windows or decaying sums~\citep{bolot2013private,upadhyay2019sublinear}. The dual approach, proposed by \citet{bolot2013private} and recently extended by \citet{andersson_continual_nodate}, instead relaxes the classical privacy definition to allow DP guarantees for past data points to gradually expire over time.

Our work is most directly inspired by the rational approximation methods of \citet{dvijotham_efficient_2024}, and in particular by their use of the equivalence between certain matrix factorizations and generating functions. While they search for simple approximations of functions to derive computationally efficient factorizations, we take the opposite approach by searching for more complicated functions to derive matrix factorizations that are usable in the unbounded setting. Because we are not directly concerned with reducing computational complexity below $O(n)$ space and $O(n\log n)$ time, we will generally take the $\mcount ^{1/2}$ mechanism  of \citet{henzinger_almost_2023} and \citet{fichtenberger2023constant} (``Sqrt Matrix'' in \autoref{tab:comparison} and \autoref{fig:variance_and_coeffs}) as the baseline against which our algorithm is compared.

\begin{table}

  \caption{Asymptotic Comparison of Private Counting Algorithms}
  \label{tab:comparison}
  \centering
      \resizebox{\textwidth}{!}{
  \begin{tabular}{@{}lcccccc@{}}
    \toprule
    \cmidrule(r){1-2}
    Mechanism & $\mathbb{V}(\mathcal{M}(X)_t) \cdot \frac{\varepsilon^2}{\log(1/\delta_{priv})
    }$ & \multicolumn{1}{p{1.4cm}}{\centering Time for $t$ outputs} & Space & Smooth & Unbounded \\    \midrule
    Binary ~\citep{chan2011private,jain_differentially_nodate}*
        & $O(\log (t)\log(n))$
        & $O(t)$
        & $O(\log t)$ 
        & No
        & No \\
    Hybrid Binary ~\citep{chan2011private}*
        & $O(\log^2 (t) + \log t)$
        & $O(t)$
        & $O(\log t)$ 
        & No
        & \textbf{Yes} \\
    Smooth Binary~\citep{andersson_smooth_nodate}      
        & $O((\log n + \log\log n)^2)$    
        & $O(t)$
        & $O(\log t)$
        & \textbf{Yes}
        & No \\
    Sqrt Matrix ~\citep{henzinger_almost_2023,fichtenberger2023constant}
        & $O(\log (t)\log(n))$
        & $O(n\log n)$
        & $O(n)$
        & \textbf{Yes}
        & No \\
    \midrule
    \textbf{\autoref{alg:log_matrix}}
        & $O(\log^{2}(t) (\log \log t)^{1+2\alpha})$
        & $O(t \log t)$
        & $O(t)$
        & \textbf{Yes}
        & \textbf{Yes} \\
    \textbf{\hyperref[par:hybrid]{Hybrid Matrix}}
        & $O(\log ^2(t) + (\log \log t)^{1+2\alpha})$  
        & $O(t \log t)$
        & $O(t)$ 
        & No
        & \textbf{Yes} \\
    \bottomrule
  \end{tabular}
  }
  \vspace{0.1em}
  
  \textit{ (*) Originally satisfied $\varepsilon$-DP with Laplacian noise; numbers reported are for the Gaussian variant. A mechanism is smooth if its variance is a smooth function of $t$, and it is unbounded if its variance at time $t$ is $O_N(1)$. Algorithms with bolded names are original to this work. Constant factors differ significantly between algorithms --- we investigate this more closely in \autoref{sec:extensions} and \autoref{fig:variance_and_coeffs}.}

\end{table}

\section{Overview of Results}

\subsection{Preliminaries and Notation}
The lower-triangular Toeplitz matrix corresponding to a sequence $a_0, a_1,\ldots$ is defined as the square matrix whose $i,j$ entry is $a_{i-j}$ when $i\geq j$ and 0 otherwise. For a given sequence, we will denote the corresponding matrix as $\ltt(a_0, a_1, \ldots)$. All matrices we consider will be infinite, but for practical computations we will work with $[A]_n$, the leading submatrix formed by the first $n$ rows and columns of $A$. The special matrix $\mcount$ is defined as $\ltt(1,1,1,\ldots)$.

$\ltt$ matrices are closed under multiplication, and $\ltt(S)\ltt(S') = \ltt(S \ast S')$, where $S \ast S'$ denotes the convolution of the sequences $S$ and $S'$. This means that multiplication of $\ltt$ matrices is commutative and can be performed in $O(n \log n)$ time using the Fast Fourier Transform (FFT). Given an analytic function $f(z)$ with Taylor coefficients $a_n = [z^n]f(z)$, we will abuse notation and write $\ltt(f)$ for $\ltt(a_0, a_1, \ldots)$. Writing $[f(z)]_n = \sum_{m=0}^n a_m z^m$ for the order-$n$ Taylor approximation of $f$, it follows that $[\ltt([f]_n)]_n = [\ltt(f)]_n$. We will also use the property that $\ltt(f)\ltt(g) = \ltt(fg)$, which follows from applying the Cauchy product to the Taylor series of $f$ and $g$. We refer the curious reader to the full version of \citet{dvijotham_efficient_2024} for an excellent and detailed introduction to $\ltt$ matrices in this context, including proofs of the properties just described.

Finally, we note that we reserve the symbol $\delta$ to refer to the exponent of the $\log\log$ term in \autoref{eq:singular}. When we need to refer to the privacy parameter of a DP algorithm, we will use $\delta_{priv}$.

\subsection{Main Result}

Our main asymptotic result is the following theorem and its associated algorithm (\autoref{alg:log_matrix}):

\begin{theorem}\label{thm:matrix}
    For all $\alpha > 0$, there exists an infinite lower-triangular Toeplitz matrix factorization $L, R \in \mathbb{R}^{\infty \times \infty}$ such that the following properties hold for all $N \leq +\infty$:
    \begin{itemize}
        \setlength\itemsep{1pt}
        \item \textbf{Joint Validity}: For all $n \in \mathbb{Z}^+$, $[L]_n[R]_n = [\mcount]_n$.
        \item \textbf{Bounded Sensitivity}: $\lim_{n \to N} \lVert [R]_n \rVert_{1\to 2} = C_N < \infty$ for some computable constant $C_N$.
        \item \textbf{Near-Optimal Asymptotic Error}:  $\lVert [L]_n \rVert_{2\to\infty}^2 = O((\log n)^2(\log \log n)^{1+2\alpha})$.
        \item \textbf{Computability}: There exists an unbounded streaming algorithm that at each time step $t$ takes as input $z_t$ and outputs $(Lz)_t$ using $O(t)$ memory and amortized $O(\log t)$ time.
    \end{itemize}
\end{theorem}

\begin{corollary}\label{cor:private-alg}
    For all $\varepsilon, \delta_{priv}, \alpha > 0$ there exists an unbounded streaming algorithm for continual counting, described in \autoref{alg:log_matrix}, which has the same complexity as in \autoref{thm:matrix} and satisfies $(\varepsilon, \delta_{priv})$-DP in the continual release model. At each time step $t$, the algorithm adds Gaussian noise with scale $O\big(\log(t)(\log\log t)^{1/2+\alpha}C_{\varepsilon, \delta}\big)$, where $C_{\varepsilon,\delta} = O(\varepsilon^{-1}\sqrt{\log(1/\delta_{priv})})$ is independent of the input.
\end{corollary}

We prove \autoref{thm:matrix} in \autoref{sec:thm1proof}; \autoref{cor:private-alg} then follows from existing results~\citep{li_matrix_nodate, denisov2022improved}.

\subsection{Technical Overview} 

The matrix decompositions we consider are of the form $L = \ltt(f(z;-\gamma, -\delta)), ~R=\ltt(f(z; \gamma, \delta))$, where:
\begin{equation}\label{eq:singular} f(z; \gamma, \delta) := f_1(z)f_2(z;\gamma)f_3(z;\delta),  \end{equation}
\begin{equation*}
    f_1(z) := \frac{1}{\sqrt{1-z}}~, \quad 
    f_2(z;\gamma) := \Big(\frac{1}{z} \log\Big(\frac{1}{1-z}\Big)\Big)^\gamma~, \quad
    f_3(z;\delta) :=  \Big(\frac{2}{z} \log\Big(\frac{1}{z} \log\Big(\frac{1}{1-z}\Big)\Big)\Big)^\delta.
\end{equation*}
We specifically consider the following parameter ranges:
\begin{equation}\label{eq:params}
    (\gamma < -1/2, \delta \geq 0) \quad \lor \quad (\gamma = -1/2, \delta < -1/2).
\end{equation}

As motivation for why one might study such matrices, recall that the first column of $\mcount$ corresponds to the Taylor coefficients of $(1-z)^{-1}$. This implies that for any pair of functions $g_1,g_2$ analytic on the open unit disc with product $(1-z)^{-1}$, we have $\mcount = \ltt(g_1)\ltt(g_2)$. Conversely, any non-pathological decomposition of $\mcount$ into LTToep matrices will correspond to such a pair of functions. So, rather than searching for arbitrary sequences of real numbers (which are complicated to represent and reason about), we can instead take the approach of searching for nice analytic functions with the hope of translating their functional properties into statements about the sequences we're ultimately interested in.

To understand why these functions in particular are good candidates, consider the special case where $\gamma=\delta=0$. This gives us $R = \ltt( (1-z)^{-1/2})$, which is exactly $\mcount ^{1/2}$ ~\citep{dvijotham_efficient_2024}. The key issue that makes this sequence inapplicable to unbounded inputs is that it is not square-summable. By Parseval's theorem, this is equivalent to the statement that the function $f$ is not square-integrable. To fix this, we would like to `nudge' $f$ so that it \textit{becomes} square-integrable while still being as `close' to the original function as possible. A plausible first attempt in this direction would be to choose $R = \ltt((1-z)^{-1/2+\alpha})$ for some $\alpha > 0$, but this gives $O(t^\alpha)$ error; the key issue is that $(1-z)^{-\alpha}$ diverges very quickly as $z \to 1^-$ even when $\alpha$ is small. To obtain logarithmic overhead, we require functions like $\log(1/(1-z))$ that diverge more slowly. 
Finally, the inclusion of the $1/z$ and $2/z$ terms in \autoref{eq:singular} eliminates the 0 of $\log(1/(1-z))$ at $z=0$, which ensures that $f$ is always analytic with $f(0)=1$. 

To study the asymptotic growth of $\lVert [L]_n \rVert_{2\to\infty}$ and $\lVert [R]_n \lVert_{1\to 2}$, we draw on the classic work of \citet{flajolet1990singularity}, whose Theorem 3B (stated formally in \autoref{thm:flajolet}) provides the following asymptotic expansion for the coefficient of $z^n$ in the Taylor series of $f$:
\[ [z^n] f(z; \gamma, \delta) \sim \frac{1}{\sqrt{n\pi}} (\log n)^\gamma (2 \log\log n)^\delta \cdot \Big(1 + O(\log^{-1}(n)) \Big). \]
In particular, this implies that if our parameters satisfy \autoref{eq:params}, then the series $\sum_{n=0}^\infty ([z^n] f(z; \gamma, \delta))^2$ converges, which is equivalent to $R$ having bounded column norm. 
This is the crucial step that allows us to avoid asymptotic dependence on $N$ without the use of doubling tricks --- by calibrating our noise scale to the partial sum $\sum_{t=0}^N ([z^t] f(z;\gamma, \delta))^2 = O_N(1)$, we can guarantee privacy for all inputs of size $n \leq N$. Using the same asymptotic expansion, we can also show that the row norm of the corresponding $L$ matrix grows like $O(\log(n)^{1/2-\gamma})$ when $\gamma < -1/2$, or $O(\log(n) (\log \log n)^{-\delta})$ when $\gamma = -1/2, \delta < -1/2$. 

Finally, we can compute the first $t$ coefficients of $f_L$ and its convolution with $z$ in $O(t \log t)$ time and $O(t)$ space using FFTs. \autoref{alg:log_matrix} combines all of these ideas together alongside a standard doubling trick on $t$, which allows us to achieve $O(n \log n)$ time complexity even when $n$ is unknown.

\begin{algorithm}[t]
    \caption{Logarithmic Matrix Method}
    \label{alg:log_matrix}
    \begin{algorithmic}
        \Require Parameters $\gamma, \delta$ satisfying \autoref{eq:params}, maximum input size $N \leq +\infty$, privacy parameter $C_{\varepsilon, \delta}$
        \State Compute $\Delta \gets \lVert [\ltt(f(z; \gamma, \delta))]_N)\rVert_{1\to 2}$\Comment{See 'Bounded Sensitivity' in \autoref{sec:thm1proof}}
        \For{$t = 1,\ldots, n$}
            \State Receive input $x_t \in [0,1]$
            \State Set $S_t \gets \sum_{i=1}^t x_i$
            \If{$t = 2^m$ for some integer $m$}
                \State Sample $z_s \sim \mathcal{N}(0, C_{\varepsilon,\delta}^2\Delta^2)$ for $t \leq s \leq 2t-1$
                \State Compute next $t$ coefficients of $L = \ltt(f(z; -\gamma, -\delta))$  in $O(t \log t)$ time
                \State \Comment{See 'Computability' in \autoref{sec:thm1proof}}
                \State Compute next $t$ terms of the sequence $Lz$ in $O(t \log t)$ time with an FFT
            \EndIf
            \State Output $S_t + (Lz)_t$
        \EndFor
    \end{algorithmic}
\end{algorithm}

\section{Proof of \autoref{thm:matrix}}\label{sec:thm1proof}

We will begin by fixing $\alpha>0$, and choosing $\gamma = -1/2, \delta = -1/2 - \alpha$. Our goal is to show that the matrix decomposition $L = \ltt(f_L), R=\ltt(f_R)$ satisfies all four properties listed in \autoref{thm:matrix}, with $f_L = f(z; -\gamma, -\delta)$, $f_R = f(z; \gamma, \delta)$, and $f$ defined as in \autoref{eq:singular}.

\paragraph{Joint Validity.} We have that $LR = \ltt(f_Lf_R) = \ltt(\frac{1}{1-z}) = \mcount$.

\paragraph{Bounded Sensitivity.} We begin by showing that the column norm of $R$ is finite, or equivalently that $f_R \in L^2$. Recall that $\lVert R \rVert_{1\to 2}^2 = \sum_{n=0}^\infty ([z^n]f_R)^2$. By Theorem 3B of \citet{flajolet1990singularity}, we know that  \( ([z^n]f_R)^2 = O(n^{-1} \log^{-1}(n)\log^{-1-2\alpha}(\log n))\), and the series therefore converges by the integral test.

To actually compute this sensitivity, we make use of Parseval's theorem, a classic result in Fourier analysis which relates square-summable sequences to functions that are square-integrable over the unit circle. Specifically, it gives us that:
\begin{equation*}
    \sum_{m=0}^\infty ([z^m] f_R)^2 = \frac{1}{2\pi} \int_0^{2\pi} |f_R(e^{i\theta})|^2~d\theta.
\end{equation*}
In \autoref{app:sensitivity}, we describe how the integral on the right-hand side can be computed numerically to high precision. We also show how this integral can be used to efficiently compute upper bounds on the partial sum $\sum_{m=0}^N ([z^m] f_R)^2$ for the case where $N < +\infty$.

\paragraph{Near-Optimal Asymptotic Error.}  Theorem 3B of \citet{flajolet1990singularity} tells us that $([z^n] f_L)^2 = O(n^{-1} (\log n) (\log\log n)^{-2\delta}) = O(n^{-1} (\log n) (\log \log n)^{1+2\alpha})$. So, by the definition of big-Oh, there are constants $n_0, C$ such that for all $n > n_0$, we compute that
\begin{align*}
    \sum_{m=n_0}^{n} ([z^m] f_L)^2 
    &\leq C \sum_{m=n_0}^{n} \frac{1}{m} (\log m)(\log\log m)^{1+2\alpha} \\
    &\leq C \int_{n_0-1}^{n} \frac{1}{x} (\log x) (\log \log x)^{1+2\alpha} ~dx \\
    &= O(\log^2(n) (\log \log n)^{1+2\alpha}),
\end{align*}
as desired.
The case where $\gamma = -1/2 - \alpha, \delta \geq 0$ can be analyzed in the same manner to obtain a bound of $O(\log^{2+2\alpha}(n))$ instead.

\paragraph{Computability} We formalize the problem we are trying to solve as a game: first, an adversary secretly fixes some integer $n > 0$. Then, at each time step $t = 1,\ldots,n$, our algorithm is required to output $(\ltt(f_L) z)_t$, where $z \sim \mathcal{N}(0, I_n)$. The algorithm's goal is to achieve an optimal asymptotic dependence on the unknown value $n$ with respect to the total memory and computation used. We will model computation in units of $M(t)$, the cost of multiplying two polynomials of degree $t$. Because polynomial multiplication can be implemented through divide-and-conquer algorithms, we assume that for any $k>0$, $M(kt) \sim k M(t)$ asymptotically as $t \to \infty$.

Initially, we disregard the challenge of not having access to $n$ and consider the intermediate problem of computing $[\ltt(f_L)]_t$ for some constant $t$ that we choose ourselves. As a preview of our eventual strategy for handling unbounded inputs, we assume that we have access to a pre-computed solution of size $t/2$. From here, we observe that $\ltt(f(z; -\gamma, -\delta)) = \ltt(f_1(z)) \ltt(f_2(z;-\gamma)) \ltt(f_3(z;-\delta))$. So, it suffices to compute each of these matrices in isolation. The final product can then be computed using $2M(t)$ (or just $M(t)$ if $\delta=0$).

To compute $\ltt(f_1(z))$, we use the fact that $f_1 = \sum_{m=0}^\infty (-z)^m {-1/2 \choose m}$. This can be used to derive the recurrence relation $[z^0] f_1 = 1, ~[z^m] f_1 = (1 - \frac{1}{2m})[z^{m-1}]f_1$ presented in prior works ~\citep{henzinger_almost_2023, fichtenberger2023constant}. This recurrence lets us compute the coefficients of this matrix in $O(t)$ time and space.

To compute the coefficients of $\ltt(f_2(z;-\gamma))$, we begin with the fact that $[z^m] f_2(z;1) = \frac{1}{m+1}$, which can be derived from the Taylor expansion of $\log(1+z)$. This intermediate sequence can also clearly be computed in linear time and space. To account for the power of $-\gamma$, we use the identity $f_2(z;-\gamma) = \exp(-\gamma\log(f_2(z;1)))$. To compute $\log(f_2(z;1))$, we use the fact that $\log'(f_2(z;1)) = \frac{f_2'(z;1)}{f_2(z;1)}$. The coefficients of $f_2'(z;1)$ can be directly computed from the coefficients of $f_2(z;1)$ through term-by-term differentiation, and using the division algorithm of \citet{hanrot2004newton} lets us compute the ratio $f_2'(z;1)/f_2(z;1)$ using $2.25 M(t)$. Term-by-term integration then allows us to recover all but the constant term of $\log(f_2(z;1))$ in linear time, but this is sufficient because we know $\log(f_2(0;1))= \log(1)=0$. 

Having shown that we can efficiently compute the Taylor coefficients of $-\gamma \log(f_2(z;1))$, it remains to find the coefficients of $f_2(z;-\gamma) = \exp(-\gamma \log(f_2(z;1))$. We achieve this with a second algorithm from ~\citep{hanrot2004newton} that takes an order $t/2$ approximation of $\exp(h)$ as input and produces an order $t$ approximation using $2.25 M(t)$. 

Finally, we can rewrite $f_3(z;-\delta)$ as $(\frac{2}{z}\log(f_2(z;1)))^{-\delta}$. This allows us to reuse our earlier computation of the Taylor coefficients of $\log(f_2(z;1))$. Dividing by $z$ corresponds to a simple shift of the coefficients. All that remains is to account for the power of $-\delta$, which can be done in $4.5M(t)$ using the same technique described above for $\gamma$. 

In total, we require $11M(t)$ to compute $[L]_t$, plus one more $M(t)$ to compute $[L]_tz$. To extend to the case of general $n$, we initialize $t=2$ and double it when we are asked to output the $(t+1)$st estimate. This gives us a total cost over the entire input of $12 \sum_{k=1}^{\lceil \log_2 n \rceil } M(2^k) \sim 12 \sum_{k=1}^{\lceil \log_2 n \rceil} 2^{-k+1} M(n) \leq 24 M(n)$.
When polynomial multiplication is implemented with FFTs, we arrive at $O(n\log n)$ time and $O(n)$ space complexity, asymptotically matching the algorithm of \citet{henzinger_almost_2023}. 
But note that because we have to adjust $t$ on the fly, our algorithm requires amortized $O(\log t)$ time at each step, compared to $O(n\log n)$ pre-processing and $O(1)$ time per step for \citet{henzinger_almost_2023}. 

\section{Implementation and Extensions}\label{sec:extensions}

In this section, we investigate various practical considerations in selecting the values of $\gamma$ and $\delta$ for \autoref{alg:log_matrix},
and propose an approximate variant (\autoref{alg:approx}) which sacrifices some theoretical rigor for a greater than $5\times$ improvement in running time for large inputs. 
Finally, we show that \autoref{alg:log_matrix} can be used as a subcomponent of the hybrid mechanism of \citet{chan2011private} to achieve exactly $O(\log^2 t)$ variance with improved constants compared to their original approach. \footnote{Our code is available at \url{https://github.com/ben-jacobsen/central-dpolo/}}

\paragraph{Computing sensitivity for finite horizons.} When $N < +\infty$, it is unnecessary to calibrate our noise to the limiting value $\lVert f_R \rVert_2^2$; the order-$N$ Taylor approximation $\lVert [f_R]_N \rVert_2^2$ suffices. The challenge is that $[f_R]_N$ typically does not have an analytic closed form, and na\"ively computing the partial sum $\sum_{n=0}^N ([z^n]f(z;\gamma, \delta))^2$ requires $\Omega(N)$ time. We seek a more efficient alternative. To that end, we introduce the function
\begin{equation}\label{eq:def-Z}
    Z(s; \gamma, \delta) \coloneq \sum_{n=0}^\infty ([z^n] f(z;\gamma, \delta))^2 e^{-sn} = \lVert f(e^{-s/2}z;\gamma, \delta) \rVert_2^2.
\end{equation}

This function has two important properties. First, it is easy to evaluate: the map $z \mapsto f(e^{-s/2}z;\gamma, \delta)$ has a simple closed form, which makes $\lVert f(e^{-s/2}z;\gamma, \delta) \rVert_2^2$ amenable to numerical integration. Secondly, $Z(s)$ is related to the Laplace transform of the cumulative sum function $P(x) = \sum_{n \leq x} ([z^n] f(z;\gamma, \delta))^2$. Specifically, define the non-negative measure $\mu = \sum_{n=0}^\infty ([z^n] f(z;\gamma, \delta))^2 \delta_n$, where $\delta_n$ denotes the Dirac delta function with argument $x-n$. Then $P(x) = \mu([0,x])$. A standard argument then yields that:
\begin{align*}
    \int_0^\infty e^{-sx} P(x)~dx &= \int_0^\infty e^{-sx} \int_{[0,x]} ~d\mu(t)~dx \\
    &= \int_0^\infty \int_t^\infty e^{-sx}~dx~d\mu(t) = \frac{Z(s)}{s},
\end{align*}
where the interchange of the integrals is justified by Tonelli's theorem. It follows that we can numerically estimate the finite-horizon sensitivity through a numerical inverse Laplace transform~\citep{stehfest1970algorithm} as
\begin{equation}\label{eq:inv-laplace}
    \lVert [f_R]_N \rVert_2^2 = \mathcal{L}^{-1} \left\{ \frac{Z(s)}{s} \right\}(N+1/2),
\end{equation}
where the offset of $1/2$ is chosen to avoid the discontinuity of $P(x)$ at integer $x$. 

\begin{figure}
    \centering
    \begin{subfigure}{0.49\linewidth}
        \includegraphics[width=\textwidth]{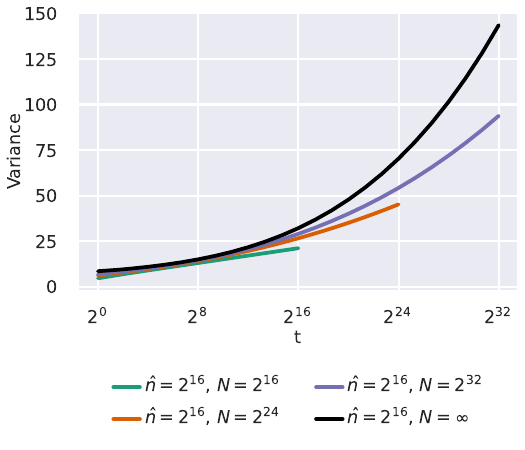}
        \caption{Fixed $\hat n$, varying $N$.}
        \label{fig:fixed-nhat}
    \end{subfigure}
    \begin{subfigure}{0.49\linewidth}
        \includegraphics[width=\textwidth]{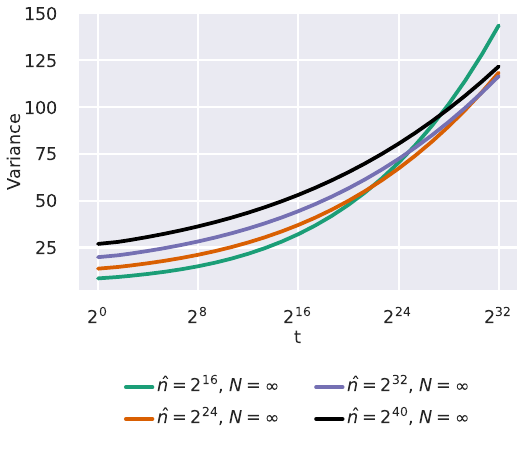}
        \caption{Fixed $N$, varying $\hat n$.}
        \label{fig:fixed-N}
    \end{subfigure}
    \caption{Illustrates the robustness of choosing matrix parameters to minimize variance at the expected input size $\hat n$ with maximum time horizon $N$. \autoref{fig:fixed-nhat}: The choice $\hat n = N = 2^{16}$ exactly corresponds to the $\mcount^{1/2}$ factorization. Choosing $N > \hat n$ leads to modestly greater error at $t \leq \hat n$ but makes it possible to process inputs that are larger than expected. \autoref{fig:fixed-N}: choosing $\hat n \gg n$ leads to uniformly increased error at all time steps, while choosing $\hat n \ll n$ gives low error initially before growing quickly once $t \gg \hat n$. It is not necessary for $\hat n$ to be a very precise estimate; at $n=2^{32}$, factorizations with $\hat n \in \{2^{24}, 2^{32}, 2^{40}\}$ achieve nearly the same maximum error despite their estimated input sizes being off by a factor of up to $256\times$. }
    \label{fig:robustness}
\end{figure}

\paragraph{Choosing the value of $\gamma$ and $\delta$.} Every value of $\gamma,\delta$ satisfying \eqref{eq:params} corresponds to a unique tradeoff between asymptotic growth rate and constant factors.
It is therefore not clear how one should identify the `best' set of parameters for a particular context. In this section, we propose a heuristic for making this choice which we show leads to good performance in practice.

As motivation, observe that if we knew the true input size $n$ but still wished to guarantee privacy for inputs of size up to $N$, then we could numerically optimize over $\gamma, \delta$ to minimize the final error term,
\begin{equation}\label{eq:param-objective}
    F(n,N) = \lVert [f_L(z)]_{n} \rVert_2^2 \cdot \lVert [f_R(z)]_N \rVert_2^2.
\end{equation}

Of course, we do not know $n$. We will resolve this issue by introducing a new parameter, $\hat n \leq N$, which represents an `educated guess' for the true input size. We expect that if this guess is at all accurate, then optimizing $F(\hat n, N)$ will also approximately optimize $F(n, N)$. We investigate the robustness of this optimization problem to misspecified $\hat n$ in \autoref{fig:robustness}.

There is also a second issue, which is that although \eqref{eq:inv-laplace} can be used to evaluate the right-hand side of \eqref{eq:param-objective}, it is not obvious \textit{a priori} that the objective is convex or otherwise easy to optimize. We resolve this second issue with the following lemma, proved in the appendix, which provides a convex proxy for our objective.
\begin{lemma}
    Define $Z(s)$ as in \eqref{eq:def-Z}. Then, for all $s \geq 0$ and $n \in \mathbb{Z}^+$, the following bound holds:
    \begin{equation*}
        U_n(s;\gamma, \delta) \coloneq e^{sn} Z(s) \geq \lVert [f(z;\gamma, \delta)]_n \rVert_2^2 = \sum_{m=0}^n ([z^m]f(z;\gamma, \delta))^2.
    \end{equation*}
    Moreover, $U_n(s;\gamma, \delta)$ is strictly convex in $s$ for fixed $\gamma, \delta$, and jointly log-convex in $\gamma, \delta$ for fixed $s$.
\end{lemma}

Because this bound holds for all $s$, it follows that $F(\hat n, N) \leq \inf_{s_L, s_R} U_{\hat n}(s_L;-\gamma, -\delta) \cdot U_N(s_R; \gamma, \delta)$. This naturally suggests an algorithmic approach similar to block coordinate descent: we can alternate between finding exact optimums over $(s_L, s_R)$ and $(\gamma, \delta)$ until the objective value converges, and this will yield a critical point of our proxy objective function.

\begin{algorithm}[t]
    \caption{Approximate Logarithmic Matrix Method}
    \label{alg:approx}
    \begin{algorithmic}
        \Require Matrix parameters $\gamma,\delta \not\in \lbrace 0, 1, \ldots \rbrace$, order $K$, threshold $\tau$ such that relative error is at most $\eta$ for $t \geq \tau$, privacy parameter $C_{\varepsilon, \delta}$, maximum input size $N \leq +\infty$
        \State Compute $\Delta \gets \lVert [\ltt(f(z;\gamma, \delta)]_N \rVert_{1 \to 2}$
        \For{$t = 1,\ldots, n$}
            \If{$t = 2^m$ for some integer $m$}
                \State Sample $z_s \sim \mathcal{N}(0, (1+\eta)^2 C_{\varepsilon,\delta}^2 \Delta^2)$ for $t \leq s \leq 2t-1$
                \If{$t < \tau$}
                    \State Use \autoref{sec:thm1proof} to compute next $t$ entries of $\ltt(f_R)$
                \Else
                    \State Use \autoref{eq:expansion} to \textit{approximate} the next $t$ entries of $\ltt(f_R)$
                \EndIf
                \State Save computed sequence as $[\hat R]_{2t}$
                \State Compute $[\hat R]_{2t}^{-1}[\mcount ]_{2t} z$
            \EndIf
            \State Output $(\mcount \hat R^{-1} z)_t$
        \EndFor
    \end{algorithmic}
\end{algorithm}

\paragraph{Constant-factor speedups through asymptotic expansions.} \citet{flajolet1990singularity} derive the following asymptotic expansion for the Taylor coefficients of $f$, which we used earlier in a weaker form to derive the asymptotic error rate of our mechanism:

\begin{theorem}[Paraphrased from \citet{flajolet1990singularity} Theorem 3B]\label{thm:flajolet}

Let $\gamma,$ and $\delta$ be complex numbers not in $\lbrace 0, 1, 2, \ldots \rbrace$. Then the Taylor coefficients of $f(z; \gamma, \delta)$ satisfy:
\begin{equation}\label{eq:expansion}
    [z^m] f(z;\gamma, \delta) \sim \frac{1}{\sqrt{m\pi}} (\log m)^\gamma (2 \log \log m)^\delta \Big(1 + \sum_{k \geq 1} \frac{e_k^{(\gamma, \delta)}(\log \log m)}{(\log (m) \log(\log (m)))^k} \Big),
\end{equation}
where $e_k^{(\gamma, \delta)}(x)$ is a polynomial of degree $k$:
\[ e_k^{(\gamma, \delta)}(x) = \sqrt{\pi} \frac{d^k}{ds^k} \big(\frac{1}{\Gamma(-s)}\big) \big\vert_{s=-1/2} \cdot E_k(x), \]
and $E_k(x)$ is the $k^{th}$ Taylor coefficient of the function $g(u)=(1-xu)^\gamma\big(1 - \frac{1}{x}\log(1-xu)\big)^\delta$.
\end{theorem}

\begin{wrapfigure}[20]{r}{2.5in}
    \centering 
    \includegraphics[]{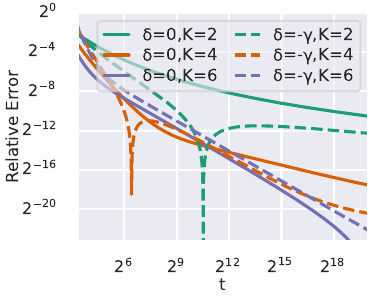}
    \captionsetup{width=2.4in}
    \caption{Plots the relative error of estimating $[z^t]f_R$ as a function of $t$ when $\gamma=-0.51$. The apparent discontinuities when $\delta > 0$ are where the additive error changes sign.}
    \label{fig:approx_error}
\end{wrapfigure}

 While this expression is messy, the cost of computing it is independent of $m$: for fixed approximation order $K$, the derivatives of the reciprocal gamma function can be precomputed and the remaining terms can be calculated in $O(K \log K)$ time using the same techniques described in \autoref{sec:thm1proof}. This idea naturally suggests \autoref{alg:approx}, which switches from exact computations to order $K$ approximations once $t$ is large enough to ensure relative error below some threshold $\eta$. We plot relative error as a function of $t, \delta$ and $K$ in \autoref{fig:approx_error}. Numerically, we find that an asymptotic order of $K=6$ is sufficient to make approximation error negligible well before the point where \autoref{alg:log_matrix} becomes computationally unwieldy.

We highlight a potential pitfall with this approach, which is that if we approximate $\hat L \approx L$ directly, then we are implicitly choosing $\hat R = \mcount \hat L^{-1}$ and can no longer calibrate our noise to $\lVert R \rVert_{1\to 2}$ as normal. Ideally we would like to be able to prove a bound like $\lVert \mcount \hat L^{-1} \rVert_{1\to2} \leq (1+O(\eta)) \lVert R \lVert_{1\to2}$, but this appears to be non-trivial. We therefore take the opposite approach, which is to directly approximate $\hat R \approx R$ and calibrate our noise to $(1+\eta)\lVert R \rVert_{1\to 2}$ with $\hat L = \hat R^{-1} \mcount$. This version of the algorithm is provably valid and private given an appropriate threshold $\tau$, and for large enough inputs it reduces the computational cost at each power of 2 from $12M(t)$ to $2.25M(t)$ (the cost of one series division), closing over $80\%$ of the computational gap between our algorithm and a standard doubling trick. The tradeoff is a multiplicative $(1+\eta)$ increase in error prior to the switch, and the loss of tight, provable bounds on error after the switch.

\paragraph{Asymptotic improvements through hybrid mechanisms.}\label{par:hybrid} In their seminal paper, \citet{chan2011private} propose a generic `Hybrid Mechanism' for continual counting of unbounded streams. The schema requires one (possibly inefficient) unbounded counting algorithm, which operates on the condensed sequence $y_k = \sum_{t=2^k}^{2^{k+1}-1} x_t$, and a bounded algorithm, which is restarted whenever $t$ is a power of 2. The privacy budget is divided between these two algorithms, and the partial sum at time $t = 2^k + r$ is given by adding together estimates for $\sum_{i=0}^{k-1} y_i$ and $\sum_{t=2^k}^{2^k+r} x_t$. The resulting hybrid mechanism is unbounded and typically preserves the exact asymptotic error rate of the chosen bounded algorithm.

\begin{figure}[t!]
    \centering
    \includegraphics[width=\linewidth]{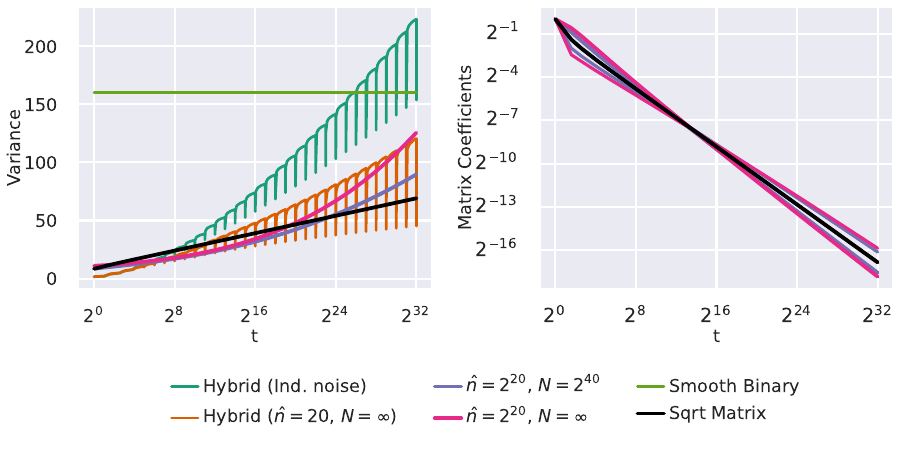}
    \caption{\textbf{Left}: Comparison of the exact variance of different algorithms and parameter choices. Contrast with asymptotics in \autoref{tab:comparison}. 
    The Hybrid mechanism with \autoref{alg:log_matrix} has significantly lower variance than the one using independent noise as in \citet{chan2011private}, but both algorithms exhibit erratic performance when $t$ is close to a power of 2. 
    For \autoref{alg:log_matrix}, calibrating sensitivity to large but finite $N$ leads to meaningfully lower error, but even the fully unbounded mechanism with $N = +\infty$ improves on the maximum error of the Hybrid mechanism for $t \leq 2^{31}$.
    All instances of the logarithmic matrix factorization have parameters chosen to numerically optimize variance at time $t = 2^{20}$; in the case of the unbounded component of the hybrid mechanisms, this corresponds to choosing $\hat n = 20$. The bounded algorithms (Sqrt Matrix and Smooth Binary) are calibrated to $n = 2^{32}$.
    \textbf{Right}: Coefficients of the matrices $\ltt(f_L)$ and $\ltt(f_R)$ for $\mcount^{1/2}$ and the two numerically optimized factorizations with $\hat n = 2^{20}, N \in \{2^{40}, +\infty\}$. The relative difference between coefficients is low, validating the heuristic that good factorizations are often `close' to $\mcount^{1/2}$.}
    \label{fig:variance_and_coeffs}
\end{figure}

The original hybrid mechanism used a simple unbounded algorithm that adds independent noise to each input (corresponding to the matrix decomposition $L = \mcount, R=I$). Our work enables a more powerful instantiation using \autoref{alg:log_matrix} as the unbounded algorithm instead (`Hybrid Matrix' in \autoref{tab:comparison}), which we compare against the original. For both implementations, we improve on the original presentation by using $\mcount^{1/2}$ for the bounded algorithm and allocating 75\% of the privacy budget to the bounded learner, which reduces the constant of the leading $O(\log^2 t)$ term. In the spirit of \citet{honaker2015efficient}, we also reuse the outputs of the bounded mechanism on earlier subsequences to optimize the variance of the final estimate. Our results are visualized in \autoref{fig:variance_and_coeffs}. 

\section{Conclusion}

We have shown for the first time that the classic matrix factorization method~\citep{li_matrix_nodate} can be efficiently extended to online settings with unknown input size \textit{without} the use of doubling tricks. The resulting algorithm is the first that we are aware of that is simultaneously \textbf{unbounded} and \textbf{smooth} with \textbf{almost-optimal asymptotic error}. Empirically, we have also shown that it enjoys excellent constants when its parameters are set correctly, with variance that is uniformly less than $2\times$ that of \citet{henzinger_almost_2023} for inputs as large as $n=2^{32}$, even when the expected input size $\hat n$ is miscalibrated by a factor of over $4000\times$.

Many interesting questions remain unanswered, however. For instance: the only algorithms we are aware of that exactly achieve asymptotically optimal $O(\log^2 t)$ variance in the unbounded setting rely on some form of doubling trick, and despite a great deal of effort, we were not able to find a $\ltt$ decomposition with this property. Our suspicion is that no such decomposition exists, and that this is related to the classic result that the series $\sum (n \cdot \log n \cdot \log\log n \cdot \ldots \cdot\log^{\circ k}n)^{-1}$ diverges for any $k$. We believe that a formal proof of this conjecture would be very interesting as it would constitute a clean separation between the power of $\ltt$ and general lower-triangular matrices in online learning. We therefore conclude by posing the following open question:

\begin{question}
    Do there exist $\ltt$ matrices $L, R$ such that $[L]_n[R]_n = [\mcount]_n$, $\lim_{n\to\infty}\lVert [R]_n \rVert_{1\to2} < \infty$, and $\lVert [L]_n \rVert^2_{2\to\infty} = \Theta(\log^2 n)$?
\end{question}

{
\small

\bibliography{OnlineLearning}

}


\newpage

\appendix

\section{Technical Details for Computing Sensitivity}\label{app:sensitivity}

To actually compute the sensitivity, we make use of Parseval's theorem, which gives us:
\begin{align*}
    &\sum_{n=0}^\infty ([z^n]f_R)^2 
    = \frac{1}{2\pi} \int_{0}^{2\pi} \lvert f_R(e^{i\theta}) \rvert^2 ~d\theta \\
    &= \frac{2^{2\delta}}{2\pi} 
    \int_0^{2\pi} 
    \left(2\sin\left(\frac{\theta}{2}\right)\right)^{-1} \left(\log^2\left(2\sin\left(\frac{\theta}{2}\right)\right) + \frac{(\pi-\theta)^2}{4} \right)^{\gamma} 
    \Bigg[ \frac{1}{4} \log^2\left(\log^2\left(2 \sin\left(\frac{\theta}{2}\right)\right) + \frac{(\pi-\theta)^2 }{4}\right) \\
    & \qquad\qquad + 
    \left(\atan2\left(\frac{\pi-\theta}{2}, -\log\left(2\sin \left(\frac{\theta}{2}\right)\right)\right) + \atan2(-\sin \theta, \cos \theta)\right)^2 ~\Bigg]^\delta ~d\theta,
\end{align*}
where $\atan2(y,x) = \arctan(y/x)$, possibly shifted by $\pm\pi$ to fall within $(-\pi, \pi]$. With the change of variables $\omega = (\pi-\theta)/2$, this simplifies to:
\begin{align*}
    \frac{2^{2\delta}}{\pi} \int_{-\pi/2}^{\pi/2} & (2\cos \omega)^{-1} \Big( \log^2(2 \cos \omega) + \omega^2\Big)^{\gamma}   \Bigg[ \frac{1}{4}\log^2\Big(\log^2(2\cos \omega) + \omega^2\Big) \\
    &+ \Big(\atan2(\omega, -\log(2\cos \omega)) + \atan2(-\sin(2\omega), -\cos(2\omega)) \Big)^2 ~\Bigg]^\delta ~d\omega
\end{align*}

From here, we define the functions:
\begin{align*}
    I_1(\omega) = 2\cos \omega \quad 
    I_2(\omega) = \log^2(I_1(\omega)) + \omega^2 \quad 
    I_3(\omega) = \frac{1}{4} \log^2(I_2(\omega)) \quad
    I_4(\omega) = \arctan\big(-\omega/\log(I_1(\omega)) + 2\omega \big).
\end{align*}
By using symmetry and splitting the interval based on the offset of $\atan2$, we finally arrive at:
\begin{align*}
    \lVert R \rVert_{1\to 2}^2 = \frac{2^{1+2\delta}}{\pi} \Bigg[ &\int_0^{\pi/3} I_1(\omega)^{-1} I_2(\omega)^{\gamma} (I_3(\omega) + I_4(\omega)^2)^\delta ~d\omega \\
    +&\int_{\pi/3}^{\pi/2} I_1(\omega)^{-1} I_2(\omega)^{\gamma} (I_3(\omega) + (I_4(\omega) - \pi)^2)^\delta~d\omega ~\Bigg].
\end{align*}

Expressed in this way, both integrands are smooth functions. It remains to determine the correct scale for numerical integration.
The integral over $[0, \pi/3]$ is unproblematic because the integrand is smooth and bounded. The second integral over $[\pi/3, \pi/2]$ is more difficult because the integrand diverges as $\omega \to \pi/2$, and a straightforward numerical quadrature will be unable to capture the extreme behavior of the tail at this scale. 
We will derive an appropriate change of variables that overcomes this issue.

Define the following auxiliary variables:

\begin{equation*} 
    \lambda(\theta) = -\log(2\sin(\theta/2)),
    \quad \omega(\theta) = (\pi-\theta)/2,
    \quad L(\theta) = \lambda^2(\theta) + \omega^2(\theta).
\end{equation*}
Then our second integral becomes:
\begin{equation*}
    \int_0^{\pi/3} \left(2\sin\left(\frac{\theta}{2}\right)\right)^{-1} L^\gamma(\theta) 
    \left( \frac{1}{4} \log^2(L(\theta)) + \left(\arctan\left( \frac{\omega(\theta)}{\lambda(\theta)}\right) - \theta \right)^2  \right)^\delta 
    ~d\theta.
\end{equation*}
Now, we introduce the change of variables $\theta = e^{-s}$, $d\theta = -\theta~ds$.
Temporarily setting aside the constant factor of $2^{2\delta}/\pi$, we are left with:
\begin{equation*}
    \int_{-\log (\pi/3)}^\infty e^{-s}
    \left(2\sin\left(\frac{e^{-s}}{2}\right)\right)^{-1} L^\gamma(e^{-s}) 
    \left( \frac{1}{4} \log^2(L(e^{-s})) + \left(\arctan\left( \frac{\omega(e^{-s})}{\lambda(e^{-s})}\right) - e^{-s} \right)^2  \right)^\delta 
    ~ds.
\end{equation*}

At this point, we divide our integral yet again at some threshold $S$, chosen so that the following asymptotic approximations hold. We can integrate over $[-\log(\pi/3), S]$ numerically since our integrand remains smooth and bounded. Beyond that region, we use the following asymptotic approximations, based on Taylor expansions:
\begin{align*}
    e^{-s}(2\sin(e^{-s}/2))^{-1} = 1+O(e^{-2s}),
    \quad \lambda(e^{-s}) = s + O(e^{-2s}),
    \quad L(e^{-s}) = s^2 + \pi^2/4 + O(e^{-s}).
\end{align*}
It follows that
\begin{equation*}
    \log L(e^{-s}) = 2\log s + \frac{\pi^2}{4s^2} + O(e^{-s}/s^2) + O(1/s^4),
\end{equation*}
and hence
\begin{equation*}
    \frac{1}{4} \log^2(L(e^{-s})) = \log^2 s \left(1 + \frac{\pi^2 }{4s^2 \log s} + O(\log s/s^4) \right).
\end{equation*}
Simultaneously, we have that
\begin{equation*}
    \frac{\omega(e^{-s})}{\lambda(e^{-s})} = \frac{\pi}{2s} + O(e^{-s}/s),
\end{equation*}
from which it follows that
\begin{align*}
    \arctan\left( \frac{\omega(e^{-s})}{\lambda(e^{-s})}\right) - e^{-s}  &= \frac{\pi}{2s} - e^{-s} - O(1/s^3) \\
    \left(\arctan\left( \frac{\omega(e^{-s})}{\lambda(e^{-s})}\right) - e^{-s} \right)^2 &= \log^2(s) \left( \frac{\pi^2}{4s^2\log^2 s} - O(1/(s^4 \log^2 s))\right).
\end{align*}

Raising these terms to the $\gamma, \delta$ powers and multiplying, we find that our integrand is equal to:
\begin{equation*}
    s^{2\gamma} \log^{2\delta}(s) \left(1 + O(1/s^2) \right).
\end{equation*}
So, to leading order, we can approximate the tail of our integral by
\begin{equation*}
    \int_S^\infty s^{2\gamma} \log^{2\delta}(s)~ds,
\end{equation*}
whose integrand has antiderivative
\begin{equation*}
    \frac{(-1-2\gamma)^{-2\delta} \Gamma(1+2\delta, (-1-2\gamma)\log (x))}{1+2\gamma},
\end{equation*}
where $\Gamma(\cdot\,,\cdot)$ is the incomplete gamma function. In the special case where $\gamma = -1/2$, this simplifies to
\begin{equation*}
    \frac{\log^{1+2\delta}(x)}{1+2\delta}.
\end{equation*}
For $\delta < -1/2$, this expression vanishes as $x \to \infty$, and so the approximate tail integral is simply equal to:
\begin{equation*}
    \frac{\log^{1+2\delta}(S)}{-1-2\delta}.
\end{equation*}

Putting all of these pieces together, we propose to compute the sensitivity of our matrix as a sum of the following three terms:
\begin{align*}
    &\frac{\pi}{2^{2\delta}} \sum_{n=0}^\infty ([z^n] f_R)^2 = \int_0^{\pi/3} 2 I_1(\omega)^{-1} I_2(\omega)^{\gamma} (I_3(\omega) + I_4(\omega)^2)^\delta ~d\omega \\
    &\quad+\int_{-\log (\pi/3)}^S e^{-s}
    \left(2\sin\left(\frac{e^{-s}}{2}\right)\right)^{-1} L^\gamma(e^{-s}) 
    \left( \frac{1}{4} \log^2(L(e^{-s})) + \left(\arctan\left( \frac{\omega(e^{-s})}{\lambda(e^{-s})}\right) - e^{-s} \right)^2  \right)^\delta 
    ~ds \\
    &\quad+\frac{(-1-2\gamma)^{-2\delta} \Gamma(1+2\delta, (-1-2\gamma)\log (S))}{-1-2\gamma}.
\end{align*}
Taking $S = 2^p$ for an appropriate value of $p$ then ensures error that is exponentially small in $p$ for the tail approximation. Meanwhile, the second integral can be numerically computed to high precision; we use the mpmath package\citep{johansson2013mpmath} for this purpose. We evaluate the stability of this calculation in \autoref{tab:tail-cutoff-stability}.

\begin{table}[t]
  \centering
  \resizebox{\linewidth}{!}{%
  \begin{tabular}{ccrrrrrrrrr}
    \toprule
    $\gamma$ & $\delta$ & $p=2$ & $p=3$ & $p=4$ & $p=5$ & $p=6$ &
    $p=7$ & $p=8$ & $p=9$ & $p=10$ \\
    \midrule
    $-0.51$ & $0$    & $\mathbf{16.5}98857$ & $\mathbf{16.5}90419$ & $\mathbf{16.58}8219$ & $\mathbf{16.587}670$ &
      $\mathbf{16.587}534$ & $\mathbf{16.587}500$ & $\mathbf{16.5874}92$ & $\mathbf{16.5874}90$ & $\mathbf{16.587489}$ \\
    $-0.51$ & $0.51$ & $\mathbf{1761.0}63888$ & $\mathbf{1761.0}61421$ & $\mathbf{1761.05}7183$ & $\mathbf{1761.05}5229$ &
      $\mathbf{1761.054}531$ & $\mathbf{1761.054}307$ & $\mathbf{1761.0542}38$ & $\mathbf{1761.05421}9$ & $\mathbf{1761.054213}$ \\
    $-0.70$ & $0.70$ & $\mathbf{10.3}55237$ & $\mathbf{10.3}52528$ & $\mathbf{10.34}8286$ & $\mathbf{10.34}6619$ &
      $\mathbf{10.34}6123$ & $\mathbf{10.3459}92$ & $\mathbf{10.3459}60$ & $\mathbf{10.34595}2$ & $\mathbf{10.345951}$ \\
    $-1.00$ & $1.00$ & $\mathbf{3.51}8784$ & $\mathbf{3.51}6067$ & $\mathbf{3.51}2372$ & $\mathbf{3.51}1243$ &
      $\mathbf{3.5109}91$ & $\mathbf{3.5109}42$ & $\mathbf{3.51093}3$ & $\mathbf{3.510932}$ & $\mathbf{3.510932}$ \\
    $-0.50$ & $-0.75$ & $\mathbf{1.18}6463$ & $\mathbf{1.18}3078$ & $\mathbf{1.182}649$ & $\mathbf{1.1825}85$ &
      $\mathbf{1.18257}5$ & $\mathbf{1.18257}3$ & $\mathbf{1.182572}$ & $\mathbf{1.182572}$ & $\mathbf{1.182572}$ \\
    $-0.50$ & $-1.50$ & $\mathbf{1.14}8382$ & $\mathbf{1.1474}72$ & $\mathbf{1.14740}8$ & $\mathbf{1.14740}2$ &
      $\mathbf{1.147401}$ & $\mathbf{1.147401}$ & $\mathbf{1.147401}$ & $\mathbf{1.147401}$ & $\mathbf{1.147401}$ \\
    \bottomrule
  \end{tabular}%
  }
  \caption{Depicts the numerical stability of our formula for $\sum_{n=0}^\infty ([z^n] f_R)^2$ as the factorization parameters $(\gamma, \delta)$ and tail
  cutoff $S=2^p$ vary. Digits are written in boldface if they match the computed value at $p=10$. For all parameter regimes we study, the estimated sensitivity quickly converges to a stable value as $p$ increases.}
  \label{tab:tail-cutoff-stability}
\end{table}

\subsection{Extension to finite N}

For $N < +\infty$, it suffices to calibrate noise to the finite horizon sensitivity $\lVert [f_R(z)]_N \rVert_2^2 = \sum_{n=0}^N ([z^n] f_R(z))^2$. 
The main challenge here is that this expression no longer corresponds to any analytic function with a closed form, and so we cannot directly compute the sensitivity with Parseval's theorem as we did in the $N=+\infty$ case. 
To circumvent this issue, let $r \in (0, 1]$, and define the function $g_r(z; \gamma, \delta) = f(rz;\gamma, \delta)$. Clearly,
\begin{equation*}
    \lVert g_r \rVert_2^2 = \sum_{n=0}^\infty r^{2n}([z^n]f(z;\gamma, \delta))^2 \geq r^{2N}\sum_{n=0}^N ([z^n] f(z; \gamma, \delta))^2.
\end{equation*}
Therefore, we have the following bound on our finite $N$ sensitivity:
\begin{equation*}
    \sum_{n=0}^N ([z^n] f(z; \gamma, \delta))^2 \leq \inf_{r\in(0,1]} r^{-2N} \lVert g_r \rVert^2_2.
\end{equation*}

The quantity $\lVert g_r \rVert_2^2$ can now be evaluated numerically using same integration methods described above - in fact, it is simpler, since there is no longer any singularity at $\theta=0$. Moreover, we claim that the infimum on the right-hand side can be found through convex optimization. To see this, let $s = -\log r$, so that our optimization problem becomes
\begin{equation*}
    \inf_{s \geq0}~\sum_{n=0}^\infty b_ne^{2s(N-n)},
\end{equation*}
where $b_n = ([z^n] f(z;\gamma, \delta))^2$. Each summand is a positive constant times the exponential of an affine function of $s$, and hence convex in $s$. In fact, it is strictly convex since all but one affine function is non-constant.
Simultaneously, we have by Parseval's theorem that:
\begin{equation*}
    \lVert g_r \rVert_2^2 = \frac{1}{2\pi} \int_0^{2\pi} |f(re^{i\theta}; \gamma, \delta)|^2~d\theta.
\end{equation*}

Recalling the definition of $f_1, f_2, f_3$ from \autoref{eq:singular}, we can rewrite
\begin{equation*}
    |f(re^{i\theta}; \gamma, \delta)|^2 = |f_1(re^{i\theta})|^2 |f_2(re^{i\theta};1)|^{2\gamma} |f_3(re^{i\theta};1)|^{2\delta} ,
\end{equation*}
which implies that
\begin{equation*}
    \lVert g_r \rVert_2^2 = \frac{1}{2\pi} \int_0^{2\pi} |f_1(re^{i\theta})|^2 \cdot \exp\left( 2\gamma \ln|f_2(re^{i\theta};1)| + 2\delta \log |f_3(re^{i\theta};1)| \right) ~d\theta.
\end{equation*}
As before, the integrand consists of a strictly positive weight times the exponential of an affine function of $(\gamma, \delta)$, which implies that $\lVert g_r \rVert$ is jointly log-convex in $(\gamma, \delta)$. We conclude that the objective function
\begin{equation*}
    U_N(s;\gamma,\delta) = e^{2sN} \lVert f(e^{-s}z; \gamma, \delta) \rVert_2^2
\end{equation*}
is log-convex in $\gamma,\delta$ (for fixed $s$) and strictly convex in $s$ (for fixed $\gamma, \delta)$. More importantly, it is an upper bound for our true objective, $\sum_{n=0}^N ([z^n] f(z;\gamma, \delta))^2$. We remark that if $N = +\infty$, the infimum over $s$ is obtained at $s=0$, which reduces to exact sensitivity in the fully unbounded setting.

\end{document}